\documentclass[12pt]{iopart}
\usepackage{iopams}  
\usepackage{lipsum}
\usepackage[]{csquotes}
\usepackage[]{units}
\usepackage{graphicx}
\usepackage{color}
\usepackage{eqnarray}
\begin{document}

\title{Phase Control of Attosecond Pulses in a Train}

\author[1]{Chen Guo$^1$, Anne Harth$^1$\footnote{The first two authors contributed equally to this work. }\footnote{Present address: Max-Planck-Institut f{\"u}r Kernphysik, Saupfercheckweg 1, 69117 Heidelberg, Germany}, Stefanos Carlstr{\"o}m$^1$, Yu-Chen Cheng$^1$, Sara Mikaelsson$^1$, Erik M\r{a}rsell$^1$\footnote{Present address: University of British Columbia - Vancouver, BC}, Christoph Heyl$^1$\footnote{Present address: JILA, NIST and the University of Colorado, 440 UCB, Boulder, CO 80309-0440, USA}, Miguel Miranda$^1$, Mathieu Gisselbrecht$^1$, Mette B. Gaarde$^2$, Kenneth J. Schafer$^2$, Anders Mikkelsen$^1$, Johan Mauritsson$^1$, Cord L. Arnold$^1$ and Anne L'Huillier$^{1}$}
\address{$^1$Department of Physics, Lund University, P. O. Box 118, SE-22100 Lund, Sweden}
\address{$^2$Louisiana State University Department of Physics and Astronomy,222-B Nicholson Hall, Tower Dr., Baton Rouge, LA 70803-4001}

\ead{Anne.LHuillier@fysik.lth.se}


\begin{abstract}
Ultrafast processes in matter can be captured and even controlled by using sequences of few-cycle optical pulses, which need to be well characterized, both in amplitude and phase. The same degree of control has not yet been achieved for few-cycle extreme ultraviolet pulses generated by high-order harmonic generation in gases, with duration in the attosecond range. Here, we show that by varying the spectral phase and carrier-envelope phase (CEP) of a high-repetition rate laser, using dispersion in glass, we achieve a high degree of control of the relative phase and CEP between consecutive attosecond pulses. The experimental results are supported by a detailed theoretical analysis based upon the semiclassical three-step model for high-order harmonic generation.  

\end{abstract}

\maketitle
\section{Introduction}

Ultrafast phenomena can be studied and even controlled by using sequences of ultrashort pulses \cite{TannorJCP1986}. This requires detailed characterization and control of the pulses, including their relative phase. The frontier in pulse duration has moved to the attosecond range using high-order harmonic generation (HHG) in gases \cite{PaulScience2001,HentschelNature2001}. However, the level of characterization and control of sequences of attosecond pulses with a central frequency in the extreme ultraviolet (XUV) spectrum and a duration reaching down to a few cycles \cite{LopezmartensPRL2005,SansoneScience2006} is far from reaching that of optical or infrared few-cycle pulses. 

The measurement of the CEP of a single attosecond pulse has been discussed theoretically \cite{HePRL2016} and recently demonstrated using high-order harmonics generated in the vacuum ultraviolet range from a solid \cite{GargN2016}. A direct measurement of the CEP in the time domain for XUV pulses is, however, so far not feasible. In contrast, the spectral phase of single attosecond pulses has been determined using cross-correlation techniques such as streaking \cite{KienbergerScience2002} with, in particular, the FROG-CRAB (Frequency-Resolved Optical Gating-Complete reconstruction of attosecond burst) analysis \cite{QuerePRL2003}. The RABBIT (Reconstruction of Attosecond Beating by Interference of Two-photon Transition) technique allows the determination of the average spectral phase of attosecond pulses in a pulse train \cite{MairesseScience2003} and is therefore well suited for multi-cycle driving pulses, such that the phase of attosecond pulses does not vary significantly between consecutive pulses, apart from the $\pi$ change, due to the fundamental symmetry of the interaction. 
The present work focuses on the {\it relative} phase change between consecutive attosecond pulses in a short train, with typically, less than five pulses, generated by a few-cycle pulse. 

The influence of the chirp of the fundamental field on the spectral width of the high harmonics has been studied previously 
\cite{ChangPRA1998,MauritssonPRA2004,KimJoPBAMaOP2004}, with the result that the spectral width of the generated harmonics becomes narrower when the fundamental field is positively chirped, due to compensation of the phase modulation due to the generation process, which leads to a negative chirp \cite{VarjuJoMO2005}. It is also well known that control of the fundamental CEP is important when HHG is driven by few-cycle pulses, since the process is sensitive to the
electric field oscillations \cite{NisoliPRL2003,SolaNP2006,HaworthNP2007}. Changing the CEP may lead to spectral shifts between odd and even orders, or for very short driving pulses, between a modulated spectrum and a quasi-continuum \cite{SolaNP2006,BaltuskaIJoSTiQE2003}. The generation of single attosecond pulses in particular requires 
precise control of the laser CEP. The study of HHG with controlled (and variable) CEP has also led to detailed study of interferences between quantum paths originating from the so-called long trajectory contributions \cite{SansonePRL2005}. 
Recently, interference effects have been observed over a broad spectral range when varying the dispersion of CEP-stable few-cycle laser pulses \cite{RudawskiTEPJD2015,HolgadoPRA2016,HolgadoPRA2017}. 

In the present work, we study HHG in argon gas as a function of chirp and CEP of a high-repetition rate CEP-stable fundamental laser field, propagating through glass with variable thickness. 
Our experimental study utilizes a state-of-the-art 200~kHz, CEP-stabilized, 6.5~fs, 850~nm laser system, based upon
optical parametric chirped-pulse amplification (OPCPA) \cite{KrebsNP2013,RudawskiTEPJD2015}. 
The excellent stability and control regarding intensity,
spectral phase and CEP of this system allows us to perform a detailed study of HHG as a function of dispersion. High-order harmonics are generated in a high-pressure gas jet, favoring the contribution of the short trajectory \cite{HeylO2016}. The harmonic spectra as a function of glass thickness, present complex interference patterns \cite{RudawskiTEPJD2015,HolgadoPRA2016} over a large (40
eV) bandwidth. To understand these structures, we
develop an analytical multiple pulse interference model, based upon
the semi-classical description of HHG \cite{CorkumPRL1993,KulanderAILF1992,LewensteinPRA1994}, which we validate by comparing with calculations
based upon the time-dependent Schr{\"o}dinger equation (TDSE) \cite{Schafer2009,Carlstroem2017}.
Combined with experimental parameters, such as precise measurements of the fundamental phase \cite{MirandaOE2012a}, our model reproduces accurately the complex interference pattern observed in the experiment, which allows us to deduce the characteristics of the underlying attosecond pulse train, including the phase difference between consecutive attosecond pulses. By finely tuning the dispersion of the fundamental field, we demonstrate control of the relative phase and CEP of consecutive pulses in a train. 

\section{Experimental method and results}

\subsection{Experimental setup}
\begin{figure}[htbp]
	\centering
	\includegraphics[]{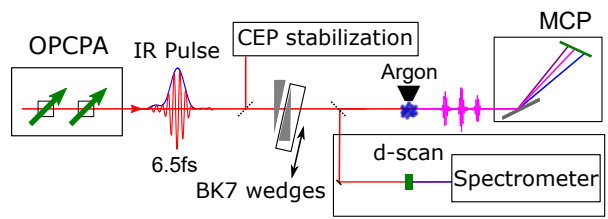}
	\caption{Experimental Setup. The dispersion of a few-cycle IR
          pulse from a CEP stabilized \unit[200]{kHz} OPCPA system is
          controlled with a BK7 glass wedge pair. The IR pulses drive
          dispersion controlled HHG in argon. With a flip mirror, the pulse can
          be characterized via a dispersion scan (d-scan) method.}
	\label{fig:setup}
\end{figure}

The laser used in our experiment is a few-cycle, 200~kHz repetition
rate, CEP stabilized OPCPA laser system \cite{Harth2017}.  The
system provides \unit[6]{$\mathrm{\mu}$J} pulses with a duration of
\unit[$<7$]{fs}. The CEP error is measured in an $f$--$2f$ interferometer to be \unit[400]{mrad} (integrated over two pulses), which corresponds to
a timing jitter of the carrier of \unit[160]{attosecond}, i.e. 12\% of one half laser cycle.  The pulse duration
is measured by a dispersion scan characterization method which uses
second harmonic generation in a thin crystal [see \fref{fig:setup}
and \cite{MirandaOE2012}].  

The laser pulses are focussed tightly, using
an achromat with a focal length of \unit[5]{cm}, into a high pressure argon
gas jet, where HHG takes place [see
\fref{fig:setup}]. The length of the medium is estimated to be
slightly larger that \unit[50]{$\mu$m} and the gas pressure right in front of the nozzle orifice to be approximately \unit[1]{bar}.  The high-pressure gas jet was designed
to optimize phase matching of the short trajectory harmonics in these
tight focussing geometrical conditions \cite{HeylO2016}.  After
passing through a \unit[200]{nm} thick Al filter in order to block the
infrared radiation (IR), the harmonics are detected by a flat-field
XUV-spectrometer, consisting of an XUV-grating and a MCP detector.
The dispersion, including obviously the CEP, of the few-cycle IR
driving pulses is varied using the same motorized BK7-glass wedge pair
that is used for the d-scan IR pulse characterization. The induced
group delay dispersion (GDD) by transmission trough BK7 is equal to
\unit[40]{fs$^2$/mm} at \unit[850]{nm}. The laser compressor,
consisting of chirped mirrors and a wedge pair, is set up in order to
precompensate transmission through air, glass (entrance window, and
achromat) such that the shortest pulse in the HHG interaction region
is obtained at the position called ``zero glass insertion''.  In order
to obtain good signal-to-noise ratios, each harmonic spectrum is
acquired by integrating over about 200~000 shots (1 second).

\subsection{Experimental results}
\begin{figure*}[!ht]
\centering
\includegraphics{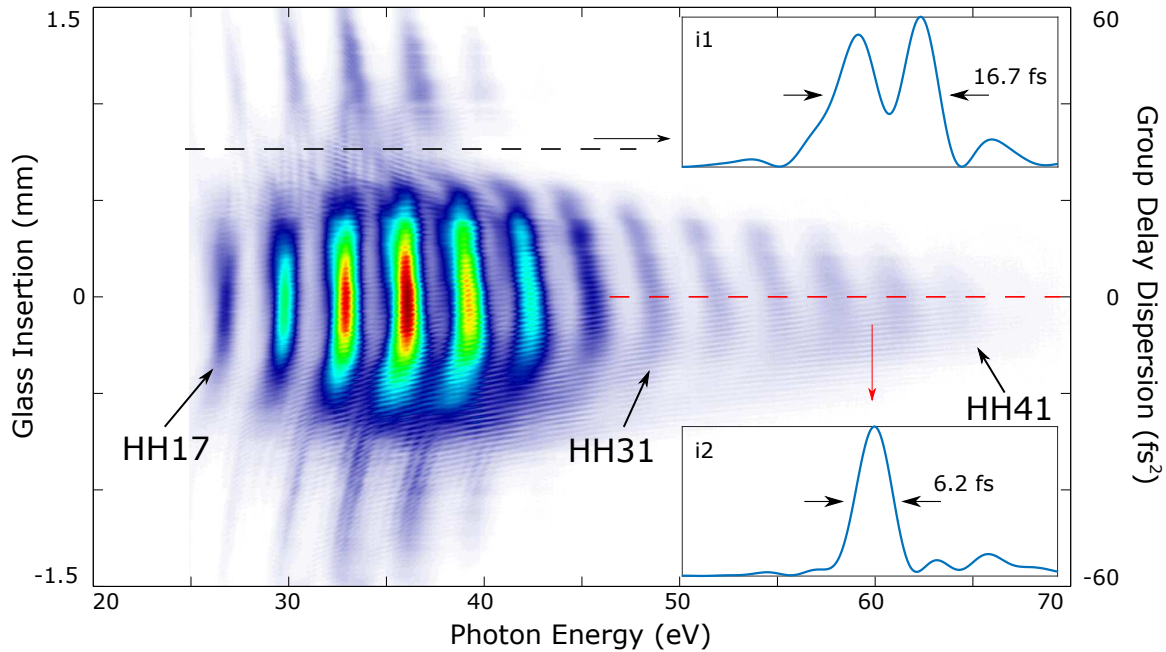}
\caption{Experimental XUV-spectra as a function of dispersion. Two inserts indicate fundamental temporal intensity profiles at i1) \unit[750]{$\mu$m}, and i2) zero glass insertion. The horizontal fringes are spaced with about $\unit[28]{\mu m}$, corresponding to a $\pi$ shift of the fundamental field CEP.}
\label{fig:exp}
\end{figure*}

The key result of this work is presented in
\fref{fig:exp} which shows the harmonic spectrum (17th
to 41st harmonic) obtained in argon gas as a function of glass insertion from
the BK7 wedge pair.  The corresponding GDD is indicated on the right
axis.
The strongest HHG signal and highest cut-off is observed for an almost 
Fourier-transform limited IR pulse at zero glass insertion.
The harmonic signal also decreases significantly for orders above the 29th or photon energy larger than \unit[45]{eV}, due to the proximity of the Cooper minimum in the
photoionization of argon, which affects the recombination step in the
single-atom response \cite{HiguetPRA2011}.
The signal decreases for large GDDs (glass insertion of $\pm$ 0.7 mm)
due to the decrease in IR-pulse intensity.
Harmonic generation can, however, be observed at large glass
insertion, an effect that we attribute to the compression of some
spectral parts of the complex IR pulse at these large dispersion
values, as retrieved from our d-scan measurements.
The harmonics, are spectrally broader for negative GDD than for
positive GDD, in agreement with previous results
\cite{KimJoPBAMaOP2004,ChangPRA1998,HolgadoPRA2016}.

In addition to the large-scale spectral features, two different interference patterns can be observed.
The most striking pattern is visible over the whole spectral range and
consists of almost horizontal fringes, separated by $\approx$
\unit[28]{$\mu$m} BK7-glass which corresponds to a $\pi$ shift of the CEP
of the driving pulse. The slope of these fringes varies from slightly
positive at negative GDD to negative at insertion values larger than
0.3 mm. As shown in more detail below, the change of slope
and asymmetry with respect to dispersion, as well as the effect on
the harmonic spectral width mentioned above, is due to the interplay
 between the chirp inherited from the fundamental spectral
 properties, and that induced by the generation process.  At a GDD
 corresponding to $\approx$ \unit[300]{$\mu$m} of glass insertion, both effects cancel each other, leading
 to spectrally narrow harmonics and horizontal CEP-fringes.
At larger insertions, around \unit[$\pm$750]{$\mu$m}, vertical
interference fringes can be observed. We attribute this effect to attosecond pulse
interferences induced by the double pulse structure of the chirped
 fundamental pulse in these conditions, as shown in the calculations presented below. 

\section{Theoretical method and results}
\label{sec:theory}

\subsection{TDSE calculations}

\begin{figure}[!ht]
	\centering
	\includegraphics{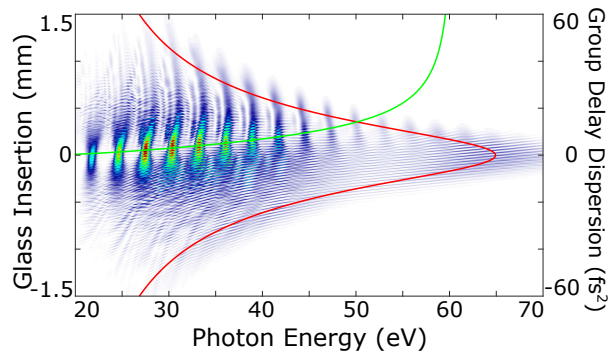}
	\caption{Calculated XUV-spectra as a function of dispersion for Gaussian pulses by solving TDSE. The red lines represent the
          position of the classical cut-off; the green curves are the calculated position where the harmonics are
          spectrally narrowed, by solving $s(\Omega)=0$ according to
          \eref{equ:general_interference1}.}
	\label{fig:scantheory}
\end{figure}

To understand our results, we first solve the TDSE in the single-active-electron
approximation \cite{Schafer2009} with an argon model atom
\cite{KulanderCPC1991}. We assume a fundamental Gaussian pulse with
\unit[6.2]{fs} pulse duration (FWHM of the temporal intensity profile)
at zero glass insertion.  The fundamental wavelength is
\unit[850]{nm}, which corresponds to the center of mass of the
experimental spectrum, and the peak intensity at Fourier-transform
limited pulse duration is \unit[2.3$\times10^{14}$]{W/cm$^2$}.  HHG spectra
are obtained by Fourier transforming the time-dependent acceleration
of the dipole moment.  We do not include propagation in the nonlinear
medium. A soft mask, which absorbs the electronic wavefunction, is placed about \unit[1.7]{nm} (\unit[32]{a.u.}) away from the nucleus. This distance is chosen using classical electron trajectory calculations so that for the shortest pulse duration, i.e.  the highest intensity, the long electron trajectories, which travel farther that the short trajectories, are absorbed, thus not contributing to the emission of radiation.  However, for lower intensity (when the glass insertion is not zero), the mask is too far away and only leads to partial absorption of the long trajectories, which therefore influence the HHG spectra.

\Fref{fig:scantheory} presents theoretical results obtained with the TDSE method.
Many of the features observed in the experiment are qualitatively reproduced.
CEP fringes are observed throughout the spectra, with a
dispersion-dependent slope. The Cooper minimum of argon is found at
about \unit[50]{eV} (31st harmonic). When the dispersion becomes positive, the harmonic peaks get narrower. 
The spectra in \fref{fig:exp} and \fref{fig:scantheory} are, however,
different at large positive or negative dispersion. The close-to-vertical fringes observed in the TDSE result cannot be explained by the distortion of the fundamental pulse [see insert in \fref{fig:exp}(i1)], as suggested for the experimental result. We believe that they might be due to the influence of the long trajectory, as discussed further below.

\subsection{Multiple interference model}
\begin{figure}[!ht]
\centering
\includegraphics{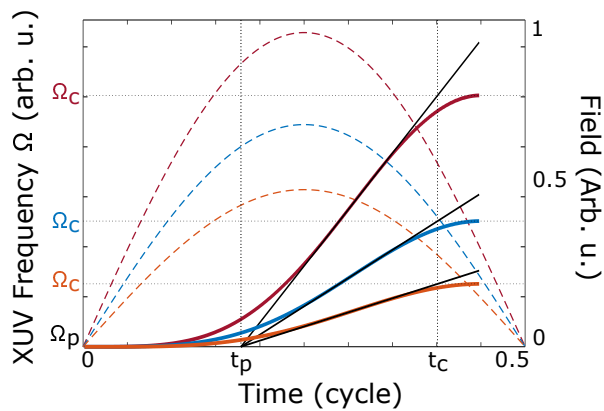}	
\caption{Fundamental electric field (dashed) and calculated emitted XUV frequency (solid) due to the short trajectory as a function of time for three intensities $I$ (red line), $I/2$ (blue line), $I/4$ (orange line). $\Omega(t)$ is obtained by solving the classical equation of motion for the electron in the field. Only one half-cycle of the fundamental is represented. $\Omega_\mathrm{p}$ is the frequency corresponding to the ionization energy $I_\mathrm{p}$, while $\Omega_\mathrm{c}$ refers to the intensity-dependent classical cut-off. $t_\mathrm{p}=0.18$ cycle and $t_\mathrm{c}=0.40$ cycle are the intersections of the tangent to the XUV frequency curve with $\Omega=\Omega_\mathrm{p}$ and $\Omega_\mathrm{c}$ respectively. }\label{fig:time}
\end{figure}

We now describe our
model, which is based upon interferences between attosecond pulses
\cite{RudawskiTEPJD2015,ManstenPRL2009}.
The attosecond light emission is described at the single atom level using the semi-classical three-step model \cite{CorkumPRL1993,SchaferPRL1993}. In this model, an electron tunnels through the potential barrier at a time $t_\mathrm{i}$, oscillates in the laser field, returns to the core at time $t_\mathrm{r}$ where it may recombine back to the ground state. The time of return $t_\mathrm{r}$ is related to $t_\mathrm{i}$ through the following equation:
\begin{equation}
 \sin(\omega t_\mathrm{r}) - \sin(\omega t_\mathrm{i})- \omega(t_\mathrm{r}-t_\mathrm{i})\cos(\omega t_\mathrm{i})=0,
 \label{return}
\end{equation} 
where $\omega$ is the laser frequency. 
The kinetic energy acquired by the electron in the field is
\begin{equation}
E_\mathrm{kin}=2U_\mathrm{p}\left[ \cos(\omega t_\mathrm{r}) - \cos(\omega t_\mathrm{i})\right]^2,
\label{kin}
\end{equation}
where $U_\mathrm{p}$ is the ponderomotive energy, equal to $e^2E_0^2/4m\omega^2$, where $e$, $m$ are the electron charge and mass and $E_0$ is the amplitude of the electromagnetic field.   
The kinetic energy reaches a maximum (a cutoff) equal to $3.2 U_\mathrm{p}$. All energies (except the cutoff) can be reached by two trajectories, the short and the long respectively. In this article, we only consider the short trajectory. 
\Fref{fig:time} shows the generated XUV frequency [$\Omega= (E_\mathrm{kin}+I_\mathrm{p})/\hbar$] as a function of return time for a half-cycle of the laser field for three different laser intensities. $\Omega_\mathrm{p}$ is the first frequency above threshold, equal to $I_\mathrm{p}/\hbar$ and $\Omega_\mathrm{c}$ is the cut-off frequency. The generated XUV frequency $\Omega$ varies approximately linearly with time of return in the HHG plateau region, as shown by comparing the exact solutions to their tangents taken at $(\Omega_\mathrm{c}+\Omega_\mathrm{p})/2$ (black lines). Remarkably, the tangent curves all cross the threshold and (intensity-dependent) cutoff frequency at the same time,    
$t_\mathrm{p}$ and $t_\mathrm{c}$ respectively. $t_\mathrm{p}$ and $t_\mathrm{c}$ are calculated numerically and found to be equal to 0.18 and 0.40 cycles of the IR laser field. The physical reason for this interesting geometrical property is that the return time is independent of intensity (\ref{return}), while the kinetic energy is proportional to it (\ref{kin}). This leads us to 
approximate the time of return $t(\Omega)=t_\mathrm{r}$ as,
\begin{equation}
t(\Omega) \approx t_{\mathrm{p}}+ \frac{t_\mathrm{c}-t_\mathrm{p}}{\Omega_\mathrm{c}-\Omega_\mathrm{p}}(\Omega-\Omega_\mathrm{p}) ,
\end{equation}
 The spectral phase $\Phi(\Omega)$ of the attosecond emission is the integral of $t(\Omega)$ so that
\begin{equation} 
\Phi(\Omega)=t_\mathrm{p}(\Omega-\Omega_\mathrm{p})+\frac{t_\mathrm{c}-t_\mathrm{p}}{\Omega_\mathrm{c}-\Omega_\mathrm{p}}\frac{(\Omega-\Omega_\mathrm{p})^2}{2},
\end{equation}
where we have dropped a constant phase term. Since $\hbar(\Omega_\mathrm{c}-\Omega_\mathrm{p})=3.2 U_\mathrm{p}$, which is proportional to the intensity $I$ within the half-cycle,
\begin{equation} 
\Phi(\Omega)=t_\mathrm{p}(\Omega-\Omega_\mathrm{p})+ \frac{\gamma}{I}(\Omega-\Omega_\mathrm{p})^2,
\label{equ:phase}
\end{equation}
where 
\begin{equation}
\gamma=\frac{\hbar(t_\mathrm{c}-t_\mathrm{p})I}{6.4 U_\mathrm{p}}=\frac{(t_\mathrm{c}-t_\mathrm{p})\hbar\varepsilon_0cm\omega^2}{3.2 e^2}.
\end{equation}
where $\varepsilon_0$ and $c$ are the vacuum permittivity and speed of light in vacuum respectively. From \fref{fig:time}, using an experimental laser cycle of \unit[2.8]{fs}, we determine $t_\mathrm{p}= \unit[0.45]{fs}$ and $\gamma=\unit[1.0\times 10^{12}]{fs^2\mathrm{W/cm}^2}$.

Equation (\ref{equ:phase}) contradicts the approximation $\Phi(\Omega)=\alpha I$, often used in the literature \cite{LewensteinPRA1995,BalcouJPB1999}. Taking the derivative of $\Phi(\Omega)$ with respect to $I$, we obtain
\begin{equation}
\alpha = \frac{\partial \Phi  }{ \partial I}= -\frac{\gamma}{I^2} (\Omega-\Omega_\mathrm{p})^2. 
\end{equation}
For a given frequency $\Omega$, $\alpha$ depends on the laser intensity. However, $\alpha$ becomes intensity-independent, if $(\Omega-\Omega_\mathrm{p}) \propto I$, i.e. if the return time is kept constant when the intensity changes. For example, for the middle point of the plateau region, $\Omega=(\Omega_\mathrm{c}+\Omega_\mathrm{p})/2$, $\alpha$ does not depend on the laser intensity since $\Omega-\Omega_\mathrm{p}= (\Omega_\mathrm{c}-\Omega_\mathrm{p})/2 \propto I$.

Our model calculates the XUV field by summing the contributions from all of the half cycles
\begin{equation}
\label{equ:general_interference0}
\tilde{A}(\Omega)=\sum_m |A_m(\Omega)| e^{i[\Omega t_m+ m\pi+\Phi_m(\Omega)] },
\end{equation}
where $m$ is the index of the half cycle of the fundamental field, with $m=0$ denoting that with the maximum amplitude, $t_m$ is the time of the zero-crossing of the electric field for the $m$-th half cycle, which corresponds to the emission time of the lowest plateau harmonic (corresponding to 0 in \fref{fig:time}). $|A_m|$ is the modulus of the spectral amplitude of the attosecond pulse emitted due to the $m$-th half cycle and $\Phi_m(\Omega)$ is the spectral phase, describing the intensity-dependent chirp of the attosecond emission [see \eref{equ:phase}]. The sign flip between consecutive attosecond pulses is described by the $m\pi$ term in the argument of the exponential.
Both CEP and dispersion of the fundamental
pulse are transferred to the attosecond pulses via the variation of the
timing $t_m$. The XUV spectrum $A_m(\Omega)$ is assumed to have a super-Gaussian shape for every attosecond pulse $m$ spanning from the threshold $\Omega_\mathrm{p}$ to the cutoff frequency $\Omega_\mathrm{c}$, which depends on the intensity of the fundamental field at $t_m$ [$I_m=I(t_m)$]. The integrated power spectrum
$\int_{0}^{\infty}|A_m(\Omega)|^2d\Omega$ of the attosecond pulses is assumed to vary with the laser
intensity as the ionization rate, which can be determined from the
Ammosov--Delone--Kra\u{\i}nov approximation \cite{AmmosovSPJ1986}. The spectral
intensity $|A_m(\Omega)|^2$ is weighted by the probability for
recombination, extracted from \cite{SamsonJoESaRP2002}. The spectral phase is obtained as explained in \eref{equ:phase} for each half cycle.

To determine the time $t_m$ and the corresponding intensity $I_m$, we calculate the field of the driving IR-pulse. We perform two calculations, using experimental and Gaussian pulses. For the experimental pulses, the spectral phase and amplitude are determined using the d-scan measurements [\fref{fig:exp}], from which the electric field for a given glass insertion $\ell$ is obtained by Fourier transform. 
The absolute fundamental CEP is not known. However, the variation of the CEP with glass insertion is included by propagating the field through glass.

For Gaussian pulses, we use an analytical formulation. We express the fundamental field as
\begin{equation}
  E(t)=E_\mathrm{max}\exp\left(-\frac{t^2}{2\tau^2}\right) 
  \sin\left(\omega t+\varphi+\frac{b}{2}t^2 \right),
    \label{equ:eoft}
\end{equation}
where $E_\mathrm{max}$ is the maximum amplitude, $\tau$ the pulse duration at $1/e$, $b$ the chirp coefficient and $\varphi$ a global phase,
 assumed to be between $-\pi/2$ and $\pi/2$. The CEP is usually defined for a cosine wave. Since we here use a sine wave, $\varphi$ is not the CEP but half $\pi/2$ shifted from it.  
The times at which the electric field goes to zero, $t_m$, are
such that
\begin{equation}
t_m =-\frac{\omega}{b} \pm \sqrt{\frac{\omega^2}{b^2}- \frac{2\varphi}{b} + \frac{2m\pi}{b}}.
\end{equation}	
Only the times with the plus sign are physically acceptable. 
The intensity for the half-cycle $m$ is given by 
\begin{equation}
I_m = I_\mathrm{max}\exp(-t_m^2/\tau^2),
\end{equation} 
where $I_\mathrm{max}$ is laser intensity at the peak of the envelope.  
Finally, we relate the chirp rate $b$, the laser intensity at the peak of the envelope $I_\mathrm{max}$ and the pulse duration $\tau$ to the glass insertion $\ell$ through the formulas:
\begin{equation}
b =\frac{2k''\ell}{a^2\tau_\mathrm{FL}^4}; I_\mathrm{max} = \frac{I_\mathrm{FL}}{a} ;
\tau=\tau_\mathrm{FL}a
\end{equation}
with
\begin{equation}
a=\sqrt{1+\frac{4k''^2\ell^2}{\tau_\mathrm{FL}^4}}.
\end{equation}
Here $\tau_\mathrm{FL}$ is the pulse duration at $1/e$ for a Fourier transform limited pulse; $k''$ is the dispersion in glass at the fundamental frequency (40.09 fs$^2$/mm); $I_\mathrm{FL}$ is the maximum laser intensity for the shortest pulse duration. 
The fundamental global phase $\varphi$ is related to the difference between phase and group velocity and is taken to be $k\ell -k'\ell \omega$, with the restriction that it should be included between $-\pi/2$ and $\pi/2$.

\begin{figure}[!ht]
	\centering
	\includegraphics{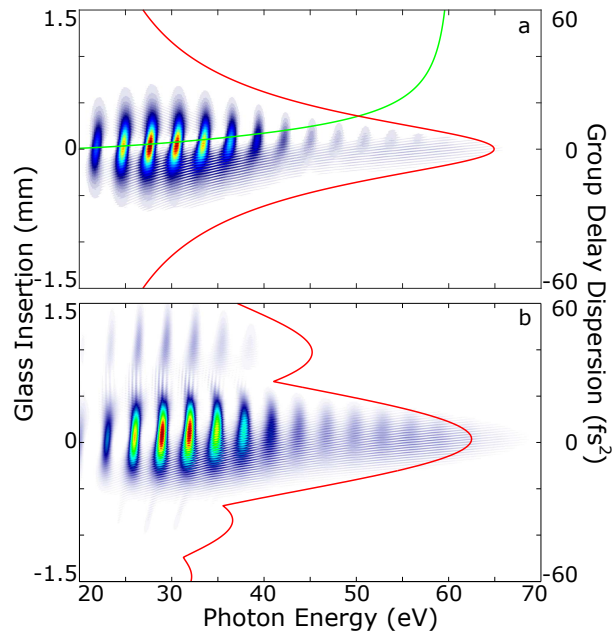}
	\caption{Calculated XUV-spectra as a function of dispersion: (a) Analytical calculation using the interference model for Gaussian pulses and (b) Numerical simulation \{from the multiple pulse interference model\} with the pulse measured from the experiment. The red lines represent the
          position of the classical cut-off; the green curves are the calculated position where the harmonics are
          spectrally narrowed, by solving $s(\Omega)=0$ according to
          \eref{equ:general_interference1}.}
	\label{fig:scanmodel}
\end{figure}

The results of the model are shown in \fref{fig:scanmodel} for Gaussian (a) and experimental pulses (b).  
Comparing \fref{fig:scantheory} and \fref{fig:scanmodel}(a), we find that most TDSE features are very well reproduced by our interference model, except for the interference fringes observed at large dispersion in the TDSE result. Since only the short trajectory contribution is included in our model, we believe that the reason for the (close-to-vertical) interference pattern in the TDSE spectra is the contribution of the long trajectories. \Fref{fig:scanmodel}(a) reproduces well the main features of the experimental spectra [\fref{fig:exp}]. In this case, the vertical fringes are due to the double pulse structure of the experimental pulse [see insert in \fref{fig:exp}(i1)].

\begin{figure}[!ht]
\centering	
\includegraphics{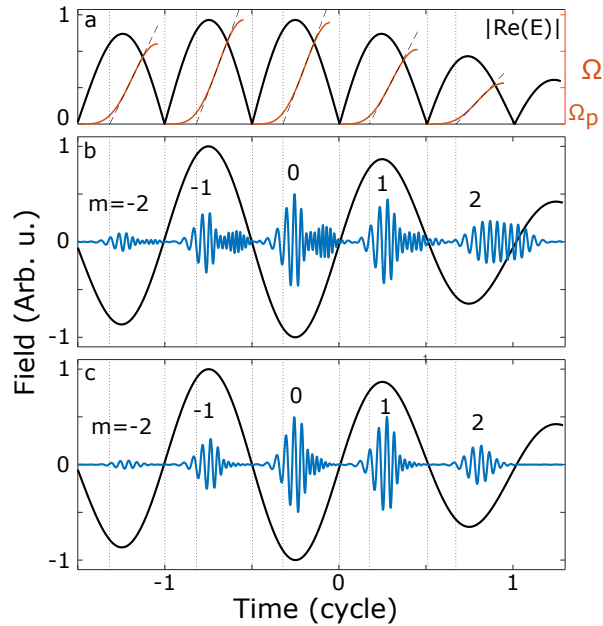}
\caption{(a)Fundamental electric field (black) and emitted XUV frequency (red) due to the short trajectory as a function of time. All of the relevant half-cycles are indicated. Generated XUV field (blue) (b) calculated using the TDSE and (c) obtained with our interference model.}\label{fig:temporal}
\end{figure}

Our model, validated by comparison with the TDSE and the experimental results, can now be used to deduce the attosecond pulse train in the time domain. \Fref{fig:temporal}(a) shows the generated XUV frequencies at each laser half-cycle during the laser pulse, while \fref{fig:temporal}(b,c) presents the attosecond pulse trains obtained at zero fundamental dispersion using TDSE and our model for a Gaussian pulse respectively. Five attosecond pulses with different chirp and timing can be
identified. Their duration varies from 220 as
at the center of the fundamental pulse to 780 as at the edges
according to the TDSE simulation. 
The three central XUV bursts in \fref{fig:temporal}(b) exhibit
a minimum in the middle. The
minimum results from the Cooper minimum at \unit[50]{eV} in the
photoionization cross-section. This spectral minimum (and the spectral phase variation associated with it, see \cite{SchounPRL2014})
is transferred to
the temporal profile of the XUV bursts through the time--energy link
inherently present in the HHG process. For the model [\fref{fig:temporal}(c)], the Cooper minimum is not as obvious as the result from TDSE, which can be attributed to a higher yield of high energy harmonics in the latter calculation. For the short trajectory
contribution, early time corresponds to low energy
\cite{SchounPRL2014}, resulting in positive chirp of the attosecond pulses \cite{MairesseScience2003}. The XUV bursts emitted before and after the three
central ones, do not exhibit this minimum since the instantaneous
intensity is too low to generate harmonics with high orders.

\subsection{Analytical derivation of the phase of the attosecond pulses} 
\label{analytic}

The excellent agreement between the TDSE calculations [\fref{fig:scantheory}] and our multiple interference model using a Gaussian pulse [\fref{fig:scanmodel}(a)], as well as between experiment [\fref{fig:exp}] and the model using experimental pulses [\fref{fig:scanmodel}(b)] motivated us to extract an approximate analytical expression for the phase of the attosecond pulses in order to understand the structure of the fringe pattern.
For small
dispersion, i.e. when $|(\varphi-m\pi) b|/\omega^2 \ll 1$, $t_m$ can be approximated by
\begin{equation}
t_m \approx -\frac{\varphi-m\pi}{\omega}-\frac{b(\varphi-m\pi)^2}{2\omega^3}.
\label{equ:tm}
\end{equation}
The phase $\Phi_m$ can be approximated by
\begin{equation}
\Phi_m(\Omega) \approx t_\mathrm{p}(\Omega-\Omega_\mathrm{p})
+\frac{\gamma}{I_\mathrm{max}}\left[1+\left(\frac{\varphi-m\pi}{\omega\tau}\right)^2\right](\Omega-\Omega_\mathrm{p})^2 
\end{equation}
where we have used $I_m \approx I_\mathrm{max}(1-t_m^2/\tau^2)$, keeping only the first term in \eref{equ:tm}. Both $t_\mathrm{p}$ and $\gamma$ do not depend on $m$.
Separating the contributions which are $m$-independent, dependent on $m$ and $m^2$, \eref{equ:general_interference0} becomes
\begin{equation}
\label{equ:general_interference1}
\tilde{A}(\Omega)=e^{iz(\Omega)}\sum_m |A_m(\Omega)| e^{imf(\Omega)+im^2s(\Omega)},
\label{equ:interfer_analytic}
\end{equation}
where the functions $z$, $f$ and $s$ are given by
\numparts
\begin{eqnarray}
	z(\Omega)=&-\frac{\Omega\varphi}{\omega}\left(1+\frac{b\varphi}{2\omega^2}\right)+
 t_\mathrm{p}(\Omega-\Omega_\mathrm{p})
	+\frac{\gamma}{I_\mathrm{max}}(\Omega-\Omega_\mathrm{p})^2 \\
f(\Omega)=&\pi+\frac{\pi\Omega}{\omega} + \varphi\kappa(\Omega) \\ 
	s(\Omega)=&-\frac{\pi\kappa(\Omega)}{2}.
\end{eqnarray}
\endnumparts
with
\begin{equation}
\kappa(\Omega)=\frac{b\pi \Omega}{\omega^3}-\frac{2 \pi\gamma }{\omega^2\tau^2I_\mathrm{max}}(\Omega-\Omega_\mathrm{p})^2.
\end{equation}
The function $z(\Omega)$ represents the phase of the ``central'' attosecond pulse in the train. The first two terms are unimportant since a linear variation in frequency leads to a shift in the temporal domain. 
The last term gives rise to group delay dispersion (GDD) which leads to temporal broadening, and which is inversely proportional to the intensity \cite{MairesseScience2003,VarjuPRL2005}. 
Furthermore, $z(\Omega)$ does not influence the spectrum $|\tilde{A}(\Omega)|^2$ and cannot be measured in our experiment. Nonlinear correlation schemes such as streaking \cite{KienbergerNature2004}, RABITT \cite{PaulScience2001} or autocorrelation \cite{TzallasNP2011} are required for characterizing attosecond pulses. 
  
The function $f(\Omega)$ describes how the CEP affects the interference between attosecond pulses, and consequently the emission at harmonic frequencies. Setting $\varphi=0$, we obtain constructive interferences when $f(\Omega)=2q\pi$, i.e. $\Omega=(2q+1)\omega$. 
The position of the constructive interferences is found to vary with the CEP through the chirp of the fundamental pulse and the dipole phase. The term $b\varphi/\omega^2$ leads to a small change in periodicity ($\delta t= \pi/\omega =T/2$ is changed into $T/2 + b\varphi T/2\omega^2$) and therefore of the frequency difference between consecutive harmonics. The dipole phase leads to a small increase of the periodicity (and therefore decrease in harmonic spacing) at high frequency.      

\begin{figure}[h]
\centering
\includegraphics{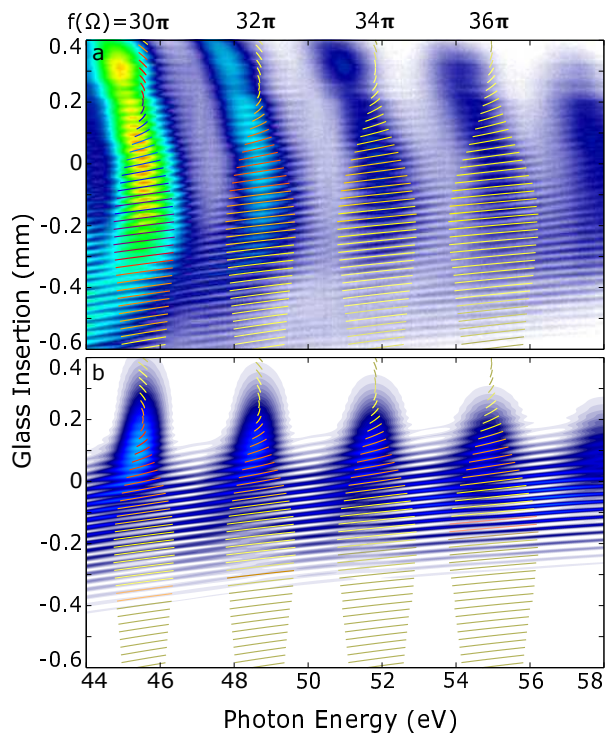}
\caption{A magnified area of (a) \fref{fig:exp} and (b) \fref{fig:scanmodel}(a), showing harmonics 31 to 37. The thin lines are the solutions to $f(\Omega)=30\pi$, $32\pi$, $34\pi$ and $36\pi$ as a function of glass insertion $\ell$. }
\label{fig:interfer}
\end{figure}

Finally, the function $s(\Omega)$ partly spoils the interference structure, leading to ``sub-harmonic'' features \cite{ManstenPRL2009}. 
The zeros of $s(\Omega)$ give the position where harmonics are sharpest. It is indicated by the green lines in \fref{fig:scantheory} and \fref{fig:scanmodel}(a) in perfect agreement with the numerical calculations. The harmonics are narrowest for positive chirp, since it compensates for the effect of the dipole phase ($\gamma>0$ for the short trajectory). Negative fundamental chirp on the other hand leads to spectral broadening of the harmonics, which eventually overlap and interfere \cite{HolgadoPRA2016}.
The function $s(\Omega)$ affects the timing between consecutive attosecond pulses, due to glass dispersion and induced by the generation process. 
Assuming $\varphi=0$ and neglecting the dipole phase, for example, the difference of time between consecutive attosecond pulses is equal to $\delta t_m= T/2- (2m-1)\pi^2 b/2\omega^3$, which increases or decreases, depending on the sign of $b$, during the laser pulse. 
Similarly, when $b=0$, the dipole phase will lead to a varying time difference between two consecutive attosecond pulses, equal to $T/2+2\pi (2m-1)\gamma(\Omega-\Omega_\mathrm{p})/\omega^2\tau^2I_\mathrm{max}$. We can make a ``perfect" train, i.e. equidistant attosecond pulses, over a certain spectral range where $s(\Omega) \approx 0$, by canceling the dipole phase variation with a small positive fundamental chirp \cite{MauritssonPRA2004}. 

Both $f(\Omega)$ and $s(\Omega)$ [through $\kappa(\Omega)$] depend on fundamental laser parameters such as chirp ($b$), pulse duration ($\tau$) and intensity ($I_\mathrm{max}$). In the limit of long pulses and no fundamental chirp, $\kappa(\Omega) \approx 0$, the pulse train becomes regular and the phase difference between consecutive attosecond pulses is equal to $\pi$ (The harmonic spectrum then consists of odd-order harmonics). 

\section{Discussion}

\subsection{Analysis of the interference fringes}

In \fref{fig:interfer}(a,b), representing magnified areas in \fref{fig:exp} and \fref{fig:scanmodel}(a) respectively, we plot the position of $f(\Omega)=2n\pi$ as a function of glass insertion $\ell$. Here, we simulated HHG with a slightly blue-shifted fundamental wavelength in order to mimic the experimental conditions. In the region $\ell=-0.6$ to +0.1 mm, the results fit well both the position of the interference fringes and their tilt with frequency as well as the dispersion-dependent width of the harmonics, which validates our model. For the trivial case of two interfering pulses, the interference pattern is governed by the function $f(\Omega)+s(\Omega)=2n\pi$. As soon as the APT includes more than two pulses, the interference pattern is essentially imposed by the condition $f(\Omega)=2n\pi$, as exemplified in \fref{fig:interfer}. Note that $f(\Omega)$ is dominated by the term $\pi\Omega/\omega$, so that $s(\Omega)$ varies with frequency much more slowly than $f(\Omega)$. For dispersion larger than $0.1$ mm, however, we believe that the interference pattern cannot only be described by the simple condition $f(\Omega)=2n\pi$ [see \eref{equ:interfer_analytic}]. 

This analysis provides a ``recipee" for retrieving the phase difference between consecutive pulses in the train, which is imprinted in the interference fringes (\fref{fig:exp}). This technique should work well for a few attosecond pulses (two or three) but becomes more complex as the number of pulses increases. An alternative method is the FROG-CRAB technique \cite{QuerePRL2003}, which, in principle, allows for the retrieval of the pulse train.   

\begin{figure*}[ht]
\centering
\includegraphics{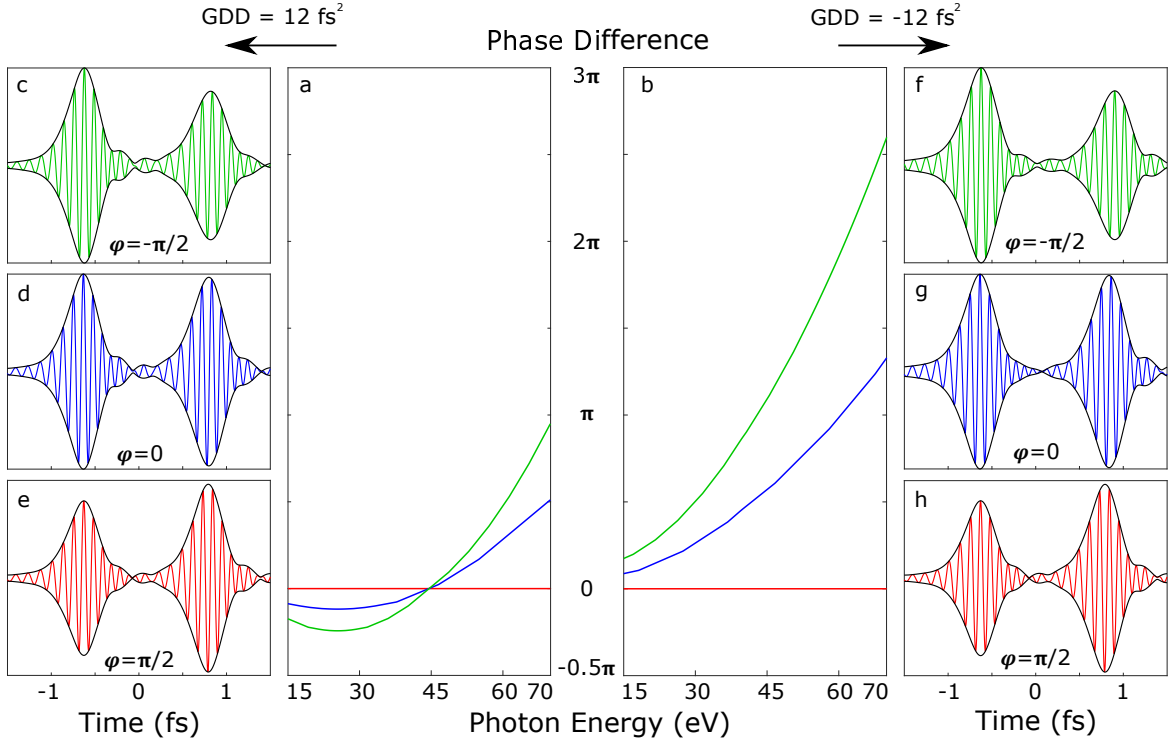}
\caption{Phase difference $\Delta\phi(\Omega)$ between the 0th and 1st attosecond pulse in the pulse train for (a) positive and (b) negative glass insertion (dispersion equal to $\unit[\pm 12]{fs^2}$, equivalent to $\unit[\pm 300]{\mu m}$ of glass insertion); Simulated 0th and 1st attosecond pulses, after spectrally filtering through a 100-nm thick chromium foil, for positive (c--e) and negative (f--h) glass insertion. The red, blue and green lines have obtained for $\varphi= \pi/2, 0$ and $-\pi/2$ respectively. An additional absolute phase is applied to the pulses to maintain the CEP of the first attosecond pulse ($m=0$) the same, for visualizing the phase variation of the second pulse ($m=1$).}
\label{fig:attophase}
\end{figure*} 

\subsection{Phase control of attosecond pulses in a train}
An important result of our derivation is that we can simply determine how the spectral phase of attosecond pulses varies from the zeroth to the first pulse (apart from the $\pi$ phase jump and the half-cycle time delay). We have
\begin{equation}
\Delta\phi(\Omega)= \left(\varphi-\frac{\pi}{2}\right) \kappa(\Omega).  
\end{equation}  
We show in \fref{fig:attophase}(a,b) this phase difference as a function of XUV photon energy for different fundamental global phase ($\varphi$) and two different glass insertions. For positive $b$, the phase difference goes through a stationary point where the effects of fundamental dispersion and dipole phase variation compensate each other [$\kappa(\Omega)=0$, green line in \fref{fig:scanmodel}(a)]. At the stationary point, the influence of $\varphi$ on $\Delta\phi$ is very small, which means that $\Delta\phi$ is very robust against any CEP fluctuations. Even when for $\varphi \neq \pi/2$, the variation of $\Delta\phi$ with $\varphi$ for any frequency remains less than $\pi$. In addition, for any fundamental CEP, the variation of $\Delta\phi$ across the spectrum is small (at most $\approx \pi$), so that, as discussed previously, the harmonics are spectrally narrow, thus leading to regular attosecond pulse trains.
In contrast, negative fundamental dispersion leads to a larger variation with CEP and across the spectrum [\fref{fig:attophase}(b)]. Here, $\Delta\phi$ can vary from 0 to almost $3\pi$ at the cutoff by changing the fundamental CEP. In this case, the harmonics are spectrally broad, and strongly CEP-dependent. The corresponding pulse trains are irregular. 

The spectral phase control demonstrated above allows us to control the relative CEP of attosecond pulses in a train. These two quantities are not independent, since the electric field is related to the complex spectral amplitude by Fourier transform. More specifically, the relative CEP of the attosecond pulse can be controlled by changing the relative spectral phase. To demonstrate this, 
 we present calculated consecutive pairs of attosecond pulses with different CEPs equal to $-\pi/2$, 0 and $\pi/2$ in \fref{fig:attophase}(c--e) and (f--h). This calculation uses the multiple pulse interference model for our experimental conditions, with the addition that a 100 nm thick chromium filter \cite{HenkeADNDT1993} is numerically introduced to select a narrow spectral range from \unit[30]{eV} to \unit[50]{eV}. A global absolute phase is also added to make the first pulse like  a ``cosine" wave (CEP equal 0), so that the phase variation of the second pulse is clearly visualized. For positive dispersion (c-d), the pulses do not change much with fundamental CEP, and the CEP difference between the two pulses is close to $\pi$. For negative dispersion (f-h), the CEP of the second pulse is equal to $\pi$ for $\varphi=\pi/2$; $\pi/2$ for $\varphi=0$ and 0 for $\varphi=-\pi/2$. Changing the fundamental CEP in this case gives us control of the relative CEP between consecutive attosecond pulses. 
The CEP control achieved by this method depends on many parameters, such as intensity, dispersion and selected spectrum.

\section{Conclusion}
 
In summary, we have studied high-order harmonic generation in argon driven by a few-cycle 200~kHz optical parametric chirped pulse amplifier system, as a function of fundamental CEP and dispersion. The spectra exhibit a complex pattern of interference fringes when the dispersion is changed.
These structures are well reproduced by simulations based on the solution of the time-dependent Schr\"odinger equation as well as by a multiple-pulse interference model, based upon the semi-classical approximation. 
Using an analytical expression for the phase of attosecond pulses in a train, we show that the relative spectral phase and CEP of consecutive
pulses in an attosecond-pulse train generated from a few cycle
CEP-stable fundamental field can be controlled by the dispersion and CEP of the driving IR pulse. Positive dispersion leads to pulse trains which are robust against fundamental CEP variation, with reproducible attosecond waveforms from one pulse to the next. In contrast, negative dispersion leads to pulse trains with variable and controllable relative atto CEP. The fundamental dispersion and CEP provide an important control knob for the attosecond pulse trains. In some applications, e.g. interferometry \cite{IsingerS2017}, robust and stable attosecond pulse trains, which can be obtained using positive dispersion, are needed. 

In other type of applications, e.g. pump/probe or coherent control \cite{TannorJCP1986}, it is important to control the relative phase between two pulses. In this case, negative dispersion and variable fundamental CEP should be used.  
 The level of control achieved in the present work extends the applicability of many coherent spectroscopy techniques, previously limited to the optical range, to shorter time scale and higher photon energy.   

\ack
This work was partly supported by the European Research Council (Grant PALP), the Marie Curie ITN MEDEA, the Knut and Alice Wallenberg Foundation, the Swedish Research Council, the Swedish Foundation for Strategic Research and Laserlab-Europe EU-H2020 654148. At LSU, this work was supported by the National Science Foundation under Grant No. PHY-1403236.

\section*{References}
\bibliography{Ref_libv2}
\bibliographystyle{iopart-num}
%

\end{document}